\documentclass[pra,preprint,showpacs]{revtex4}

\usepackage{amsmath}
\usepackage{graphicx}
\usepackage[T1]{fontenc}

\newcommand{\avg}[1]{\langle #1\rangle}
\newcommand{\ket}[1]{| #1\rangle}
\newcommand{\bra}[1]{\langle #1|}

\begin{document}

\title{Statistics of multiphoton events in spontaneous parametric down-conversion}
\bibliographystyle{apsrev}
\author{Wojciech Wasilewski}\email{wwasil@fuw.edu.pl}
\affiliation{Institute of Experimental Physics, Warsaw University, Ho{\.z}a 69, 00-681 Warsaw, Poland}
\author{Czes{\l}aw Radzewicz}
\affiliation{Institute of Experimental Physics, Warsaw University, Ho{\.z}a 69, 00-681 Warsaw, Poland}
\author{Robert Frankowski and Konrad Banaszek}
\affiliation{Institute of Physics, Nicolaus Copernicus University,
Grudzi\k{a}dzka 5, 87-100 Torun, Poland}

\begin{abstract}
We present an experimental characterization of the statistics of multiple photon pairs produced by spontaneous
parametric down-conversion realized in a nonlinear medium pumped by  high-energy ultrashort pulses from a regenerative amplifier. The photon number resolved measurement has been implemented with the help of a fiber loop detector. We introduce an effective theoretical description of the observed statistics based on parameters that can be assigned direct physical interpretation. These parameters, determined for our source from the collected experimental data, characterize the usefulness of down-conversion sources in multiphoton interference schemes that underlie protocols for quantum information processing and communication.
\end{abstract}

\maketitle

\section{Introduction}

Spontaneous parametric down-conversion (SPDC) is the basic source of non-classical light in experimental quantum optics \cite{HOMI}, testing foundations of the quantum theory \cite{BellSPDC}, and implementing protocols for quantum information information processing and communication \cite{SPDCQIPC}. The essential feature of SPDC is the guarantee that the photons are always produced in pairs, and suitable arrangements allow one to generate various types of classical and quantum correlations within those pairs.

\begin{figure}
    \begin{center}
    \includegraphics[scale=0.25]{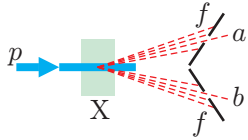}
   \caption{(Color online) A schematic of a spontaneous parametric down-conversion source.
   The nonlinear crystal $X$ is pumped with a laser beam $p$. Generated photons
   are highly correlated and useful modes $a$ and $b$ are typically selected by narrow spatial and frequency
   filters $f$.}
   \label{fig:source}
    \end{center}
\end{figure}

The physics of SPDC depends strongly on optical properties of nonlinear media in which the process is realized. This leads to an interplay between different characteristics of the source and usually imposes trade-offs on its performance. For example, many experiments require photon pairs to be prepared in well-defined single spatio-temporal modes. In contrast, photons generated in typical media diverge into large solid angles and are often correlated in space and time, as shown schematically in Fig.~\ref{fig:source}. Specific modes can be selected afterwards by coupling the output light into single-mode fibers and inserting narrowband spectral filters. However, it is usually not guaranteed that both the photons in a pair will always have the matching modal characteristics, and in many cases only one of the twin photons will get coupled in
\cite{URen}. This effect, which can be modelled as a loss mechanism for the produced light, destroys perfect correlations in the numbers of twin photons. These losses come in addition to imperfect detection, and can be described jointly using overall efficiency parameters.

The effects of losses become more critical when the SPDC source is pumped with powers so high that it is no longer possible to neglect the contribution of events when multiple pairs have been simultaneously produced \cite{RarityNJP06}. Such a regime is necessary to carry out multiphoton interference experiments, it can be also approached when increasing the production rate of photon pairs. One is then usually interested in postselecting through photocounting the down-conversion term with a fixed number of photon pairs and observing its particular quantum statistical features \cite{SunLiu}.
In the presence of losses the same number of photocounts can be generated by higher-order terms when some of the photons escape detection. However, the statistical properties of such events can be completely different, thus masking the features of interest. Although some quantum properties may persist even in this regime, with a notable example of polarization entanglement \cite{DurkSimoPRA04}, their extraction and utilization becomes correspondingly more difficult.

The present paper is an experimental study of multiphoton events in spontaneous parametric down-conversion with particular attention paid to the effects of filtering and losses. The multiple-pair regime is achieved by pumping the nonlinear crystal by the frequency-doubled output of a 300~kHz titanium-sapphire regenerative amplifier system. The kilohertz repetition rate has allowed us to count the number of the photons at the output with the help of the loop detector \cite{Loopy}.
Using a simplified theoretical description of the SPDC source we introduce effective parameters that characterize
its performance in multiphoton experiments. The obtained results illustrate trade-offs involved in experiments with multiple photon pairs and enable one to select the optimal operation regime for specific applications.

This paper is organized as follows. First we describe a theoretical model for SPDC statistics in Sec.~\ref{Sec:SPDCstat}. Sec.~\ref{Sec:Parameters} introduces effective parameters to characterize SPDC sources. The experimental setup and measurement results are presented in Sec.~\ref{Sec:Exp}. Finally, Sec.~\ref{Sec:Conclusions} concludes the paper.

\section{SPDC statistics}
\label{Sec:SPDCstat}
We will start with a simple illustration of the effects of higher-order terms in SPDC. Suppose for simplicity that the source produces a two-mode squeezed state which can be written in the perturbative expansion as $\ket{0,0}+r\ket{1,1}+r^2\ket{2,2}+\ldots$, where $r$ measures squeezing and is assumed to be real. For two-photon experiments, the relevant term is $\ket{1,1}$ and the contribution of
the higher photon number terms can be neglected as long as $r \ll 1$. This enables postselecting
the two-photon term and observing associated quantum effects, such as Hong-Ou-Mandel interference. Suppose now that
each of the modes is subject to losses characterized by $1-\eta$, where $\eta$ is the overall efficiency. Losses may transform the term $\ket{2,2}$ into $\ket{2,0}$ or $\ket{0,2}$, whose presence will lower the visibility of the Hong-Ou-Mandel interference. The two-photon term now occurs with the probability $\eta^2 r^2$, while the four-photon term effectively produces one of the states $\ket{2,0}$ or $\ket{0,2}$ with the total probability equal to $2(1-\eta)^2 \eta^2 r^4$. This constitutes a fraction of $2(1-\eta)^2 r^2$ of the events that come from single pairs produced by the source. This fraction can easily become comparable with one, especially when the losses are large.

Let us now develop a general model of photon statistics produced by an SPDC source. In the limit of a classical undepleted pump the output field is described by a pure multimode squeezed state. By a suitable choice of spatio-temporal modes, called characteristic modes, such a state can be brought to the  normal form \cite{NormalForm} in which modes are squeezed pairwise. Denoting the annihilation operators of the characteristic modes by $\hat{a}_k$ and $\hat{b}_k$, the non-vanishing second-order moments can be written as:
\begin{align}\label{Eq:<aa>,<ab>}
\avg{\hat a_k^\dagger \hat a_{l}}&=\avg{\hat b_k^\dagger \hat b_{l}}
=  \frac{\delta_{kl}}{2} (\cosh 2 r_k -1 )\nonumber \\
\avg{\hat a_k \hat b_{l}}&=\avg{\hat a_k^\dagger  \hat b_{l}^\dagger }^* =
\frac{\delta_{kl}}{2}\sinh 2 r_k  &
\end{align}
where $r_k$ is the squeezing parameter for the $k$th pair of modes. Because the state of light produced in SPDC is gaussian, these equations, combined with the fact that first-order moments vanish $\langle \hat{a}_k \rangle
= \langle \hat{b}_k \rangle = 0$ define fully quantum statistical properties of the output field.

Let us first consider the case when the spatial and spectral filters placed after the source select effectively
single field modes. The annihilation operators $\hat{a}$ and $\hat{b}$ are given by linear combinations of the
characteristic modes:
\begin{align}\label{Eq:cd_observed}
\hat a=&\sum_k t_k \hat a_k & \hat b=&\sum_k t'_k \hat b_k
\end{align}
where $t_k$ and $t'_k$ describe amplitude transmissivities of the filters for the characteristic modes
and $\sum_k |t_k|^2 = \sum_k |t'_k|^2 = 1$. Because the complete multimode state is gaussian, the reduced
state of the modes $\hat{a}$ and $\hat{b}$ is also gaussian, and it is fully characterized by the average
numbers of photons $\bar n=\avg{\hat a^\dagger \hat a}$ and $\bar n'=\avg{\hat b^\dagger \hat b}$, and
the moment $S=\avg{\hat a \hat b}$. These quantities can be written in terms of the multimode moments given
in Eq.~(\ref{Eq:<aa>,<ab>}), but we will not need here explicit expressions. It will be convenient
to describe the quantum state of the modes $\hat{a}$ and $\hat{b}$ with the help of the Wigner function:
\begin{equation}
W_{\hat{\rho}}(\alpha,\beta)=\frac{1}{\pi^2\sqrt{\det \mathbf C}}
\exp\left(-\frac{1}{2}
\boldsymbol{\alpha}^\dagger \mathbf{C}^{-1}
\boldsymbol{\alpha}\right)
\label{Eq:Wmix=gaussian}\end{equation}
where $\boldsymbol{\alpha}=(\alpha,\alpha^\ast,\beta,\beta^\ast)^T$ and
$\mathbf C$ is the correlation matrix composed of symmetrically
ordered second order moments:
\begin{equation}
\mathbf C
=
\begin{pmatrix}
\bar{n}+\frac{1}{2} & 0 & 0 & S \\
0 & \bar{n}+\frac{1}{2} & S^\ast & 0 \\
0 & S & \bar{n}'+\frac{1}{2} & 0 \\
S^\ast & 0 & 0 & \bar{n}'+\frac{1}{2}.
\end{pmatrix}.
\end{equation}

Given the Wigner function of the reduced state for the modes $\hat{a}$ and $\hat{b}$ in the Gaussian form, the calculation of the joint count statistics $\rho_{n,m}$ is straightforward. It will be useful to introduce an operator representing the generating function of the joint count statistics
\begin{equation}
\hat \Xi(x,y)=\sum_{n,m} x^n y^m\ket{n,m}\bra{n,m} = x^{\hat{a}^\dagger\hat{a}} y^{\hat{b}^\dagger \hat{b}}
\end{equation}
whose expectation value over the quantum state expanded into the power series yields the joint count statistics
$\rho_{nm}$:
\begin{equation}
\rho_{n,m} = \left. \frac{1}{n!m!} \frac{d^{n+m}}{dx^n dy^m} \avg{\hat{\Xi}(x,y)} \right|_{x=y=0}
\end{equation}
Because the operator $\hat \Xi(x,y)$ is formally equal, up to a normalization constant, to product
of density matrices
describing thermal states of modes $\hat{a}$ and $\hat{b}$ with average photon numbers $x/(1-x)$ and
$y/(1-y)$ respectively, the corresponding Wigner function has a Gaussian form:
\begin{multline}
W_{\hat \Xi}(\alpha,\beta)= \frac{4}{\pi^2(1+x)(1+y)} \\
\times\exp\left(-2|\alpha|^2\,\frac{1-x}{1+x}\right) \exp\left(-2|\beta|^2\,\frac{1-y}{1+y}\right).
\end{multline}

Using the above expression it is easy to evaluate the generating function for the
joint count statistics by integrating the product of the respective Wigner functions:
\begin{align}
\nonumber
\avg{\hat \Xi(x,y)} & =\pi^2 \int d^2\alpha \,
d^2\beta\, W_{\hat \Xi}(\alpha,\beta) W_{\hat{\rho}}(\alpha,\beta)
\\
\label{Eq:Xi(x,y)=squeezedwlosses}
 & =\frac{1}{N+1-N(\eta x+1-\eta)(\eta' y+1- \eta')}
\end{align}
where we introduced the following three parameters:
\begin{align}\label{Eq:r2,eta=onemode}
\eta &=\frac{1}{\bar n'}(|S|^2-\bar n \bar n') \nonumber \\
\eta'&=\frac{1}{\bar n}(|S|^2-\bar n \bar n') \\
N&=\frac{\bar n \bar n'}{|S|^2-\bar n \bar n'} = \frac{\bar n}{\eta} = \frac{\bar n'}{\eta'} \nonumber
\end{align}
These three parameters have a transparent physical interpretation in the regime when
$|\avg{\hat a \hat b}|^2>\avg{\hat a^\dagger \hat a}\avg{\hat b^\dagger \hat b}$. Then, it is easy to check that
the generating function given in Eq.~(\ref{Eq:Xi(x,y)=squeezedwlosses}) describes the count statistics
of a plain two-mode squeezed state whose two modes have been sent through lossy channels with transmissivities
$\eta$ and $\eta'$, and the average number of photons produced in each of the modes was equal to $N$. Note that the inequality $|\avg{\hat a \hat b}|^2>\avg{\hat a^\dagger \hat a}\avg{\hat b^\dagger \hat b}$ implies nonclassical
correlations between the modes $\hat{a}$ and $\hat{b}$. If the opposite condition is satisfied, the state can be represented as a statistical mixture of coherent states in modes $\hat{a}$ and $\hat{b}$ with correlated amplitudes. In this case, it is not possible to carry out an absolute measurement of losses.

\section{Parameters}
\label{Sec:Parameters}

In a realistic situation, the spectral filters employed in the setup are never sufficiently narrowband to ensure completely coherent filtering. Therefore a sum of counts originating from multiple modes will be observed. We will model this effect by assuming that the detected light is composed of a certain number of $\mathcal M$ modes
with identical quantum statistical properties described in the preceding section. The generating function $\Xi(x,y)$ for count statistics is therefore given by the $\mathcal M$-fold product of the expectation value $\avg{\hat{\Xi}(x,y)}$ calculated in Eq.~(\ref{Eq:Xi(x,y)=squeezedwlosses}):
\begin{equation}
\Xi(x,y) = \avg{\hat{\Xi}(x,y)}^{\mathcal M}
\end{equation}
The parameter $\mathcal M$, which we will call the equivalent number of modes, can be read out from the variance of the count statistics in one of the arms characterized by the generating function $\Xi(x,y)$.
For a single mode source the variance is equal to that of a thermal state with $(\Delta n)^2=\avg{n}+\avg{n}^2$.
It is easily seen that for $\mathcal M$ equally populated modes the variance becomes reduced to the value $(\Delta n)^2=\avg{n}+\avg{n}^2/\mathcal M$. Solving this relation for $\mathcal M$ leads us to a measurable parameter that will help us to characterize the effective number of detected modes:
\begin{equation}
\mathcal M=\frac{\avg{n}^2}{(\Delta n)^2-\avg{n}}.
\end{equation}
Note that $\mathcal M$ is closely related to the inverse of the Mandel parameter \cite{MandelQ}.

The second parameter we will use to characterize the SPDC source measures the overall losses experienced by the produced photons. Let us first note that in the perturbative regime, when all the squeezing parameters $r_k \ll 1$,
each one of the quantities $\bar n$, $\bar n'$, and $|S|^2$ appearing in Eq.~(\ref{Eq:r2,eta=onemode})
is proportional to the pump intensity. In this regime the efficiencies can be approximated by
the ratios $\eta \approx |S|^2/\bar n'$ and $\eta' \approx |S|^2/\bar n$ that are independent of the pump intensity.
On the other hand, the average photon numbers $\avg n$ and $\avg {n'}$ in both the arms are linear in the pump power.

Following early works on squeezing \cite{RaymerPRL_69PRL2650}, we will introduce here a parameter
that quantifies the subpoissonian character of
correlations between the counts $n$ and $n'$ in the two arms of the setup. First we define a stochastic variable:
\begin{equation}
 \delta= \left(\cfrac{ n}{\avg{n}}-\cfrac{n'}{\avg{n'}}\right)\Bigg/
{\sqrt{\cfrac{1}{\avg{ n}}+\cfrac{1}{\avg{ n'}}}}\,.
\end{equation}
This definition takes into account the possibility of different losses in the two arms through a
suitable normalization
of the count numbers. The subpoissonian character of the correlations can be tested by measuring the average
$\avg{\delta^2}$. The semiclassical theory predicts
$\avg{\delta^2}\ge 1$, while for two beams with count statistics characterized by the generating function
$\avg{\hat\Xi(x,y)}^{\cal M}$ one obtains:
\begin{equation}
\avg{\delta^2}=1-2\Big/\left(\frac{1}{\eta}+\frac{1}{\eta'}\right) < 1.
\end{equation}
In the case of equal efficiencies $\eta=\eta'$ this expression reduces simply to $\avg{\delta^2}=1-\eta$.
We will use the last relation as a method to measure the average overall efficiency of
detecting the state produced by the SPDC source.

\begin{figure}
    \begin{center}
    \includegraphics[scale=0.9]{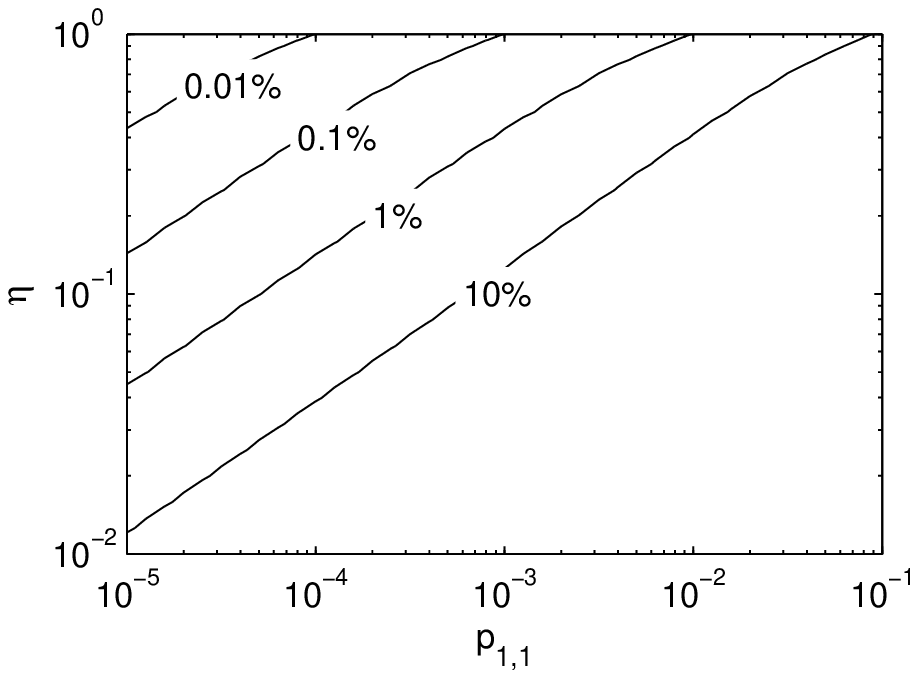}
    \includegraphics[scale=0.9]{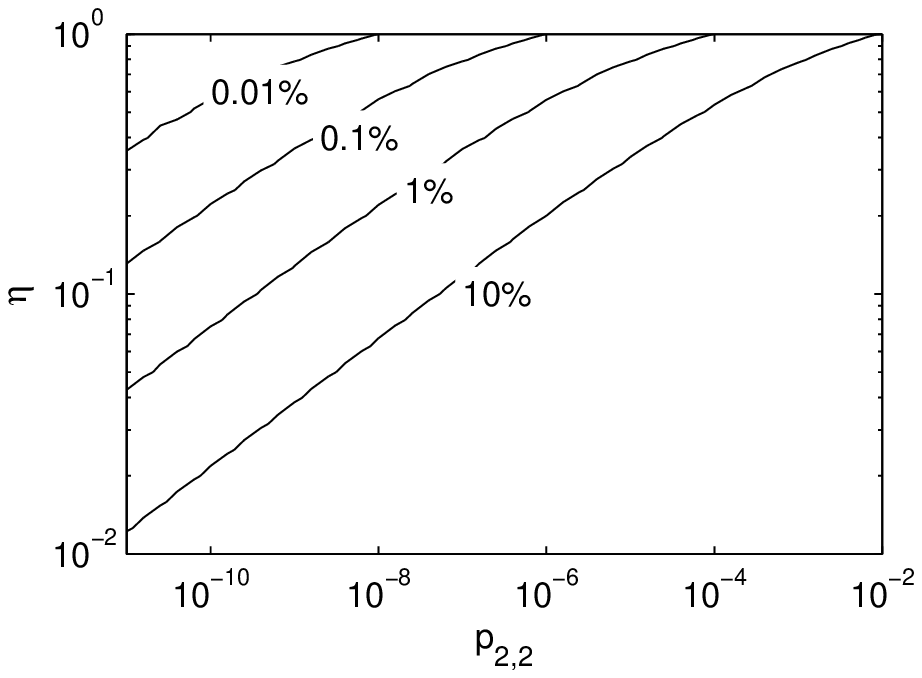}
   \caption{Contour plots of the contamination parameters for single photon pairs $\varepsilon_2$ (upper plot) and
   double photon pairs $\varepsilon_2$ (lower plot) as a function of the overall efficiency $\eta$ and the respective
   production rates $p_{1,1}$ and $p_{2,2}$. The calculations have been carried out 
   in the regime of single selected modes when $\mathcal{M}=1$.}
   \label{fig:2imp}
    \end{center}
\end{figure}

As discussed at the beginning of Sec.~\ref{Sec:SPDCstat}, the initial quantum state of light used for
many experiments should ideally be in a state $\ket{1,1}$ or $\ket{2,2}$. In the perfect case of two-mode squeezing
and negligible losses, such states can be isolated through postselection of the SPDC output on the appropriate number
of counts. In practice, the postselected events will also include other combinations of input photon numbers. Because typical detectors have limited or no photon number resolution, one needs to take into account also the deleterious contribution of higher total photon numbers. This leads us to the following definition of the parameter
measuring the contamination of photon pairs with other terms that cannot be in general removed through postselection:
\begin{equation}
\varepsilon_2 = 1-\frac{\rho_{1,1}}{\sum_{k+l\ge2} \rho_{k,l}}
\end{equation}
An analogous definition can be given for quadruples of photons, which ideally should be prepared in a state $\ket{2,2}$:
\begin{equation}
\varepsilon_4 = 1-\frac{\rho_{2,2}}{\sum_{k+l\ge4} \rho_{k,l}}.
\end{equation}
In Fig.~\ref{fig:2imp} we depict contour plots of the contamination parameters $\varepsilon_2$ and
$\varepsilon_4$ as a function of the production rates and the overall efficiency. It is clearly seen that the non-unit efficiency imposes severe bounds on the production rates that guarantee single or double photon pair events sufficiently free from spurious terms. The graphs also imply that strong pumping is not a sufficient condition to achieve high production rates, but it needs to be combined with a high efficiency of collecting and detecting photons.

\section{Experimental results}
\label{Sec:Exp}
\
\begin{figure}
    \begin{center}
    \includegraphics[scale=0.65]{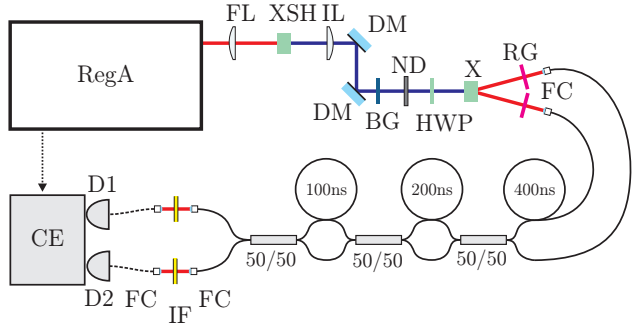}
   \caption{(Color online) Experimental setup. RegA, Regenerative amplifier; FL, IL, lenses;
   XSH, second harmonic crystal; DM, dichroic mirrors; BG, blue glass filter;
   ND, neutral density filter; HWP, half wave plate; X, downconversion crystal;
   RG, red glass filters; FC, fiber coupling stages; IF, interference filter;
   D1, D2, avalanche photodiodes; CE, coincidence electronics. Thin solid lines
   indicate single mode fibers, dashed lines --- multimode fibers.}
   \label{fig:setup}
    \end{center}
\end{figure}
The experimental setup is depicted in the Fig.~\ref{fig:setup}. The master laser (RegA 9000 from Coherent)
produces a train of 165~fs FWHM long pulses at a 300~kHz repetition rate centered at the wavelength 774~nm,
with
300~mW average power. The pulses are doubled in the second harmonic generator XSH
based on a 1~mm thick beta-barium borate (BBO)
crystal cut for a type-I process. Ultraviolet pulses produced this way have 1.3~nm bandwidth and 30~mW average
power. They are filtered out of the fundamental using a pair of dichroic mirrors DM and a color glass filter BG
(Schott BG39), and imaged using a 20~cm focal length lens IL on a downcoversion crystal X, where they form a spot
measured to be 155~$\mu$m in diameter. The power of the ultraviolet pulses, prepared in the polarization perpendicular to the plane of the setup, is adjusted using neutral
density filters ND and a motorized half waveplate. The type-I down-conversion process takes place in
a 1~mm thick BBO crystal X cut at
$29.7^\circ$ to the optic axis, and oriented  for the maximum source intensity.

The down-converted light emerging at the angle of $2.5^\circ$ to the pump beam is coupled into a pair of single mode fibers placed at the opposite ends of the down-conversion cone. The fibers and the coupling optics define
the spatial modes in which the down-conversion is observed \cite{FiberModes}. The coupled photons enter the loop
detector \cite{Loopy} in which light from either arm can propagate towards one of the detectors through eight
distinct paths. The minimal delay difference between two paths is 100~ns, more than twice the dead time of the
detectors. Finally the photons exit the fiber circuit, go through interference filters IF and are coupled
into multimode fibers which route them directly to single photon counting modules SPCM (PerkinElmer
SPCM-AQR-14-FC) connected to fast coincidence counting electronics (custom-programmed Virtex4 protype board
ML403 from Xilinx) detecting events in a proper temporal relation to the master laser pulses. The measurement
series are carried out with pairs of interference filters of varying spectral widths.

\begin{figure}
    \begin{center}
    \includegraphics[scale=0.6]{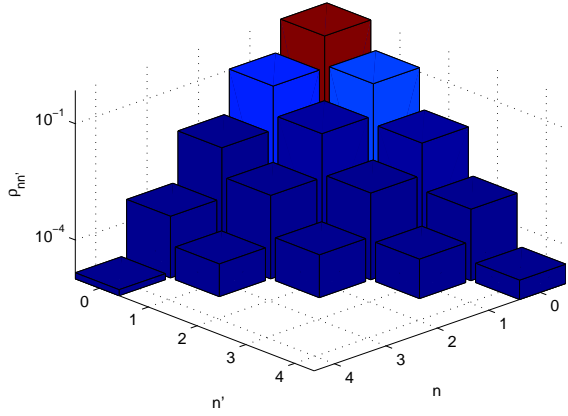}
   \caption{An exemplary joint photon number distribution measured without spectral filters.
   The average photon numbers in the two arms are $\langle n \rangle $=0.15 and $\langle n'\rangle$=0.18.}
   \label{fig:exprho}
    \end{center}
\end{figure}

The measurement proceeds as follows: for each pair of interference filters the loop detector is calibrated
using data collected at a very low pump light intensities, when the chance of more than one photon entering
the fiber circuit is negligible compared to the rate single photons appear. This allows one to calculate the complete matrix of conditional probabilities $P_{k,n}$ of observing $k$ detector clicks with $n$ initial
photons \cite{Loopy,PaulPRL96} for one arm,
and analogously $P'_{l,m}$ for the second arm. The losses in the detectors and in the fiber circuit are assumed to contribute to the overall efficiencies $\eta$ and $\eta'$. Thus $P_{k,n}$ and $P'_{l,m}$ describe lossless loop detectors for which $P_{1,1}=1$. After the calibration, the counts
are collected for approximately $10^8$ master laser pulses for each chosen intensity of the pump, which yields the probabilities $p_{k,l}$ of
observing $k$ clicks in one arm and $l$ in the other one. These probabilities are related to the joint count probability $\rho_{n,m}$ corrected for combinatorial inefficiencies of the loop detector through the formula:
\begin{equation}
p_{k,l} = \sum_{n,m} P_{k,n} P'_{l,m} \rho_{n,m}.
\end{equation}
The probabilities $\rho_{n,m}$ can be retrieved from experimentally measured $p_{k,l}$ using the maximum likelihood estimation technique \cite{MaxLik}. An exemplary joint photon number distribution reconstructed from the experimental data is shown in Fig.~\ref{fig:exprho}. The results of the reconstruction are subsequently used to calculate the parameters of the source discussed in the preceding section.

\begin{figure}
    \begin{center}
    \includegraphics[scale=0.9]{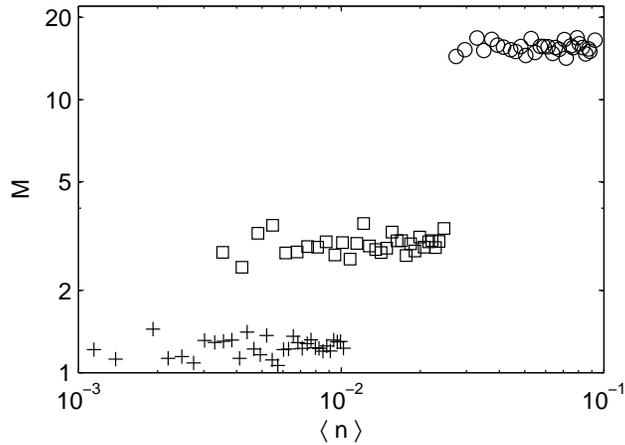}
   \caption{The equivalent number of modes $\mathcal M$ in a single arm as a function of the average photon number
   $\langle n \rangle$, for measurements carried out
   without interference filters (circles), with 10~nm FWHM
   filters (squares) and 5~nm FWHM filters (crosses).}
   \label{fig:M}
    \end{center}
\end{figure}

In Fig.~\ref{fig:M} we depict the reconstructed
equivalent number of modes $\mathcal M$ in single arm for different filtering and pump intensities.
This quantity can be also understood as a measure of how many incoherent modes
a photon from the source occupies. Naturally, $\mathcal M$ drops with application of narrowband
spectral filtering since it
erases the information on the exact time the photon pair was born in the nonlinear crystal. It also is seen that
$\mathcal M$ is practically independent on the pumping intensity, as predicted in Sec.~\ref{Sec:Parameters}.

\begin{figure}
    \begin{center}
    \includegraphics[scale=0.9]{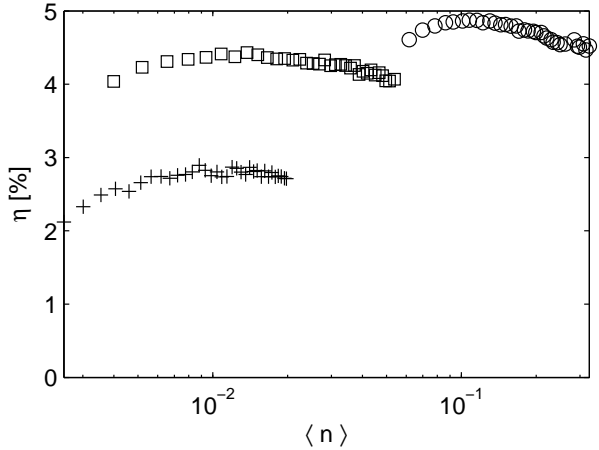}
   \caption{The equivalent detection efficiency $\eta$ as a function of the total average photon number $\langle n \rangle$, measured without interference filters (circles), with 10~nm FWHM
   filters (squares) and 5~nm FWHM filters (crosses).}
   \label{fig:eta}
    \end{center}
\end{figure}
The average overall efficiency of the squeezed state calculated as $\eta = 1 - \avg{\delta^2}$
is shown in Fig.~\ref{fig:eta}. It is
seen that the efficiency decreases with an application of narrowband filtering, which again is easily understood. In addition, $\eta$ exhibits a very weak dependence on the pump intensity which again agrees with
theoretical predictions for the regime when the average number of photons is much less than one.
A relatively large difference in $\eta$ between 10~nm and 5~nm interference filters can be explained by the fact that in the latter case the selected bandwidth becomes narrower than the characteristic scale of spectral correlations within a pair. Consequently,
the filter in one arm selects only a fraction of photons conjugate to those that have passed through the filter placed
in the second arm. This observation is consistent with the determination of the parameter ${\cal M}$, which shows that for 5~nm filters effectively single spectral modes are selected.

\begin{figure}
    \begin{center}
    \includegraphics[scale=0.9]{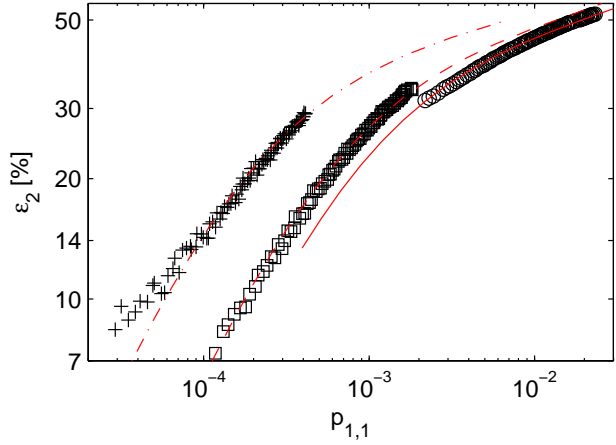}
   \caption{(color online) The contamination parameter $\varepsilon_2$ for single photon pairs as a function of the pair production rate,
   measured without interference filters (circles), with 10\/nm FWHM
   filters (squares) and 5\/nm FWHM filters (crosses).
   The parameters of the fitted theoretical curves are: $\mathcal{M}=16.6$ and $\eta=4.5\%$ (red solid line),
   $\mathcal{M}=3.1$ and $\eta=4.2\%$ (red dashed line)
   and $\mathcal{M}=1.5$ and $\eta=2.6\%$ (red dash-dot line).
   }
   \label{fig:2impexp}
    \end{center}
\end{figure}

\begin{figure}
    \begin{center}
    \includegraphics[scale=0.9]{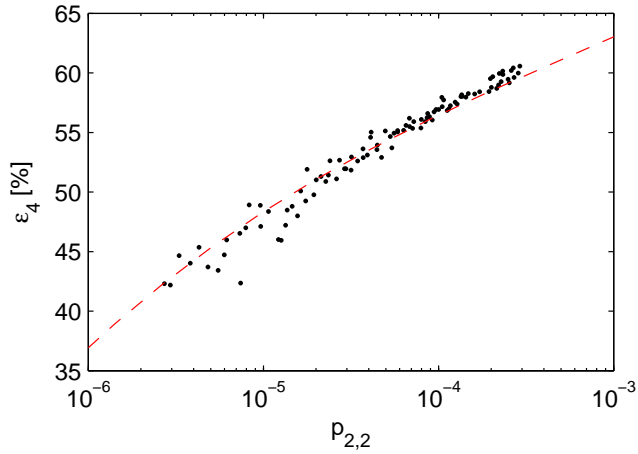}
   \caption{(color online)
   The contamination coefficient $\varepsilon_4$ for double photon pairs as a function of the production
   rate, measured without interference filters (dots). The parameters of the fitted theoretical curve are:
   $\mathcal{M}=16$ and $\eta=4.9\%$ (red dashed line).
   Measurements with interference filters did not provide statistically
   significant data.}
   \label{fig:4impexp}
    \end{center}
\end{figure}

Finally, in Figs.~\ref{fig:2impexp} and \ref{fig:4impexp} we plot the respective contamination parameters for single $\varepsilon_2$ and double $\varepsilon_4$ photon pairs, that
describes non-postselectable contributions of other photon terms to the output. As expected, the contamination grows with decreasing overall efficiency $\eta$ that corresponds to narrowing the spectral bandwidth, as well as
with the increasing pair production rates. It is noteworthy that
the measurement results agree well with fitted theoretical curves whose parameters match those that can be read out from Figs.~\ref{fig:M} and \ref{fig:eta}.
It is seen that in the regime the experiment was carried out the contamination is rather significant, and it would affect substantially effects such as two-photon interference.
This limits in practice the energy of pump pulses---and consequently the production rates---that result in photon pairs sufficiently free from spurious terms. The effect is particularly dramatic in the case of $\varepsilon_4$, where genuine double pairs appear only in approximately half of all the events.

\section{Conclusions}
\label{Sec:Conclusions}

Despite the simplicity of the basic concept, the application of SPDC sources in more complex quantum optics and quantum information processing experiments requires a careful choice of operating conditions. The actual output state is a result of a subtle interplay between the pump strength, spatial and spectral filtering of the output, and the losses experienced by the signal. In this paper we concentrated on the features of the photon number distribution and developed an effective theoretical description. Starting from the fact that the SPDC output processed by arbitrary passive linear optics is given by a gaussian state, we introduced effective parameters characterizing the joint photon number distribution. The first parameter is the effective number of orthogonal modes that impinge on each of the detectors. This parameter carries information about the modal purity of the photons produced in each arm. The second parameter characterizes average losses experienced by the SPDC output, and can be calculated as a suitably normalized variance of the count difference. Finally, we also proposed and determined experimentally a measure of the contamination of photon pairs by other spurious terms that in general cannot be rejected by postselection and may contribute to an unwanted background. We showed that the presence of such a background puts stringent requirements on the detection efficiency if a bright source with high pair generation probability is desired. These constraints become very challenging when designing multiple-pair experiments.

\begin{acknowledgments}
We acknowledge helpful discussions with C. Silberhorn and I. A. Walmsley.
This work has been supported by the Polish budget funds for
scientific research projects in years 2005-2008 and the European Commission under the Integrated Project Qubit
Applications (QAP) funded by the IST directorate as Contract Number 015848. The experiment has been carried out in the National Laboratory for Atomic, Molecular, and Optical Physics in Toru\'{n}, Poland.
\end{acknowledgments}


\begin{thebibliography}{18}

\bibitem{HOMI}
C. K. Hong, Z. Y Ou, and L. Mandel. Phys. Rev. Lett. {\bf 59}, 2044 (1987).

\bibitem{BellSPDC}
P. G. Kwiat, E. Waks, A. G. White, I. Appelbaum, and P. H. Eberhard, Phys. Rev. A
{\bf 60}, R773 (1999);
T. Paterek, A. Fedrizzi, S. Gr\"{o}blacher, T. Jennewein, M. \.Zukowski,
M. Aspelmeyer, and A. Zeilinger, Phys. Rev. Lett. {\bf 99}, 210406 (2007);
C. Branciard, A. Ling, N. Gisin, C. Kurtsiefer, A. Lamas-Linares, and V. Scarani,
{\em ibid.}\/ {\bf 99}, 210407 (2007).

\bibitem{SPDCQIPC}
J. L. O'Brien, G. J. Pryde, A. G. White, T. C. Ralph, D. Branning,
Nature {\bf 426}, 264 (2003);
P. Walther, K. J. Resch, T. Rudolph, E. Schenck, H. Weinfurter, V. Vedral, M. Aspelmeyer,  and A. Zeilinger,
{\em ibid.} {\bf 434}, 169 (2005);
M. Halder, A. Beveratos, N. Gisin, V. Scarani, C. Simon, and H. Zbinden,
Nature Physics {\bf 3}, 692 (2007);
G. Vallone, E. Pomarico, F. De Martini, and P. Mataloni,
Phys. Rev. Lett. {\bf 100}, 160502 (2008).

\bibitem{URen}
A.B. U'Ren, K. Banaszek, I.A. Walmsley, Quantum Inf. Comput. {\bf 3}, 480 (2003).

\bibitem{RarityNJP06}
O. Alibart, J. Fulconis, G. K. L. Wong, S. G. Murdoch, W. J. Wadsworth, and J. G. Rarity,
New J. Phys. {\bf 8}, 67 (2006).

\bibitem{SunLiu}
F. W. Sun, B. H. Liu, Y. F. Huang, Z. Y. Ou, and G. C. Guo,
Phys. Rev. A {\bf 74}, 033812 (2006).

\bibitem{DurkSimoPRA04}
G. A. Durkin, C. Simon, J. Eisert, and D. Bouwmeester, Phys. Rev. A {\bf 70}, 062305 (2004);
H. S. Eisenberg, G. Khoury, G. A. Durkin, C. Simon, and D. Bouwmeester, Phys. Rev. Lett. {\bf 93},
193901 (2004).

\bibitem{Loopy}
D. Achilles, C. Silberhorn, C. \'{S}liwa, K. Banaszek, and I. A. Walmsley,
Opt. Lett. {\bf 28}, 2387 (2003); M. J. Fitch, B. C. Jacobs, T. B. Pittman, and J. D. Franson,
Phys. Rev. A {\bf 68}, 043814 (2003).

\bibitem{NormalForm}
Arvind, B. Dutta, N. Mukunda, and R. Simon, Pramana-J. Phys. {\bf 45}, 471 (1995);
R. S. Bennink and R. W. Boyd, Phys. Rev. A {\bf 66}, 053815 (2002);
A. Botero and B. Reznik, {\em ibid.} {\bf 67}, 052311 (2003);
S. L. Braunstein, {\em ibid.} {\bf 71}, 055801 (2005);
W. Wasilewski, A. I. Lvovsky, K. Banaszek, and C. Radzewicz, {\em ibid.} {\bf 73}, 063819 (2006).

\bibitem{MandelQ}
L. Mandel, Opt. Lett. {\bf 4}, 205 (1979).

\bibitem{RaymerPRL_69PRL2650}
D. T. Smithey, M. Beck, M. Belsley, and M. G. Raymer, Phys. Rev. Lett. {\bf 69}, 2650 (1992).

\bibitem{FiberModes}
F. A. Bovino, P. Varisco, A. M. Colla, G. Castagnoli, G. Di Giuseppe, and A. V. Sergienko,
Opt. Comm. {\bf 227}, 343 (2003);
S. Castelletto, I. P. Degiovanni, A. Migdall, and M. Ware
New J. Phys. {\bf 6}, 87 (2004);
A. Dragan, Phys. Rev. A {\bf 70}, 053814 (2004).

\bibitem{PaulPRL96}
H. Paul, P. T\"{o}rm\"{a}, T. Kiss, and I. Jex,
Phys. Rev. Lett. {\bf 76}, 2464 (1996).

\bibitem{MaxLik}
Z. Hradil, Phys. Rev. A {\bf 55}, R1561 (1997);
D. Mogilevtsev, Opt. Comm. {\bf 156}, 307 (1998);
K. Banaszek, G. M. D'Ariano, M. G. Paris, and M. F. Sacchi,
Phys. Rev. A 61, 010304 (2000).

\end{thebibliography}
\end{document}